\documentclass[aps,longbibliography,preprint]{revtex4-1}
\usepackage{amsfonts,amssymb,amsmath,bm}
\usepackage[dvips]{graphicx}
\usepackage{color}
\usepackage[dvipsnames]{xcolor}
\usepackage{hyperref}
\usepackage{times}
\usepackage{rotating}

\definecolor{aqua}{rgb}{0.0, 1.0, 1.0}
\definecolor{bostonuniversityred}{rgb}{0.8, 0.0, 0.0}
\definecolor{dukeblue}{rgb}{0.0, 0.0, 0.61}

\begin{document}

\title{Virtual wave stress in deep-water crossed surface waves}

\author{Vladimir M. Parfenyev}
\author{Sergey S. Vergeles}\email{ssver@itp.ac.ru}

\affiliation{Landau Institute for Theoretical Physics, Russian Academy of Sciences, 1-A Akademika Semenova av., 142432 Chernogolovka, Russia}
\affiliation{National Research University Higher School of Economics, Faculty of Physics, Myasnitskaya 20, 101000 Moscow, Russia}

\date{\today}

\begin{abstract}
Waves excited on the surface of deep water decay in time and/or space due to the fluid viscosity, and the momentum associated with the wave motion is transferred from the waves to Eulerian slow currents by the action of the virtual wave stress. Here, based on the conservation of the total momentum, we found the virtual wave stress produced by calm gravity waves under the assumption that the slow Eulerian currents are weak in the sense that the Froude number is small and the scattering of surface waves by slow flow inhomogeneities can be neglected. In particular, we calculated the virtual wave stress generated by a propagating wave and two orthogonal standing waves. The obtained results possess Euler invariance, are consistent with previously known ones, and generalize them to the case of the excitation of almost monochromatic waves propagating in arbitrary directions.
\end{abstract}

\maketitle

\section{Introduction}

It was shown by Stokes that in a surface wave excited in an ideal fluid the Lagrangian particles possess a second-order drift velocity, which is usually called the Stokes drift \cite{stokes1847theory}. Later, Longuet-Higgins found that the fluid viscosity substantially modifies the Stokes prediction \cite{longuet1953mass}. The correction is associated with Eulerian slow current that corresponds to the mean velocity of fluid inside the bulk. The Stokes drift and the Eulerian slow current are very different. The Stokes drift is the result of nonlinear Lagrangian dynamics during one time period of oscillations and it does not produce any contribution into the mean velocity of the fluid (in the Euler description), while the slow current is excited by a force, which is localized in the narrow viscous sublayer near the fluid surface and is produced due to hydrodynamic nonlinearity (it is also known as the virtual wave stress; see Ref.~\cite{longuet1969nonlinear}). Therefore, the dynamics of the slow current is relatively slow and it is determined by the fluid viscosity and inertia. In the stationary regime, the slow current is independent of fluid viscosity, even though it originates from the viscosity.

The origin of the force which excites the Eulerian slow current is similar to that of the force which produces the acoustic streaming in fluid during the propagation of a sound wave \cite{boluriaan2003acoustic}. Both forces are of the second-order in the wave amplitude and linear in the fluid viscosity. The acoustic streaming finds numerous applications in microfluidics \cite{friend2011microscale,wiklund2012acoustofluidics}, as it enables remote flow excitation and object manipulation.
The standard theoretical approach to derive an equation describing the acoustic streaming flow is to average hydrodynamic equations over fast wave oscillations. The acoustic flow is excited near the boundaries inside the viscous sublayer, where the viscosity reduces the amplitude of the sound wave. Thus, the approach needs to resolve the viscous sublayer which is parametrically thinner compared to the typical spatial scale of the acoustic flow.

The same applies to the existing theoretical treatment of the excitation of a slow current by waves propagating on the free surface of a fluid \cite{longuet1953mass, filatov2016nonlinear}. Nevertheless, there is an important difference between these phenomena. The boundaries that confine the fluid in acoustic experiments produce stresses acting on the fluid, whereas there is no external force acting on the free surface of the fluid. Thus, in the latter case, one can reformulate the hydrodynamic equations in the form of conservation laws and then utilize them to find the virtual wave stress that excites the current \cite{longuet1969nonlinear, weber2001virtual}.

In this paper, we propose the treatment of the virtual wave stress produced by calm gravity waves in the deep water approximation based on the momentum conservation law, which binds together the damping of surface waves due to the viscous dissipation and the nonlinear generation of the mean fluid velocity. Roughly saying, the momentum associated with the wave motion decreases together with the wave amplitude during the propagation of the surface wave due to the fluid viscosity. However, the total momentum must be conserved and it means that the wave attenuation gives rise to the force, which excites the additional fluid flow. We demonstrate that this force is applied near the fluid surface within the crest-trough layer, which includes the oscillating boundary of the fluid and the viscous sublayer under the boundary. The thickness of the crest-trough layer is small as compared to the scale of the slow current, so the force can be treated as surface stress. We obtain the explicit expression for it in terms of the excited wave motion. Our approach avoids the fine resolution of the viscous sublayer and allows us to obtain simple equations describing the slow currents under the assumption that they are weak in the sense that the Froude number is small enough and the scattering of surface waves by slow flow inhomogeneities can be neglected. In the case of a progressive wave excited on the surface of deep water, we can reproduce the Longuet-Higgins' result \cite{longuet1953mass}.

The initial interest in this problem was inspired by the recently observed phenomenon of nonlinear vorticity generation by crossed surface waves \cite{filatov2016nonlinear, francois2017wave}. Using the established general expression for the virtual wave stress, we check that its curl corresponds to the boundary condition for the vertical vorticity used in Ref.~\cite{filatov2016nonlinear}. On the whole, the developed approach allows one to look at the generation of slow currents by crossed surface waves from a new angle, reveals the physical nature of this phenomenon, and generalizes the previous results to the case of the excitation of almost monochromatic waves propagating in arbitrary directions.

\section{Problem Formulation}

We consider an incompressible flow with a free surface that corresponds to the surface gravity waves excited against a background of a slow current. The wave motion has a characteristic frequency $\omega$ and its spectral width $\Delta \omega$ is assumed to be small, $\Delta \omega \ll \omega$. The axis $Z$ is directed vertically, opposite to the gravitational acceleration $\bm g$, and the fluid surface is determined by the equation $z=h(t,x,y)$ (it coincides with the plane $z=0$ at rest). The wave breaking is absent, the wave steepness is small, $|\nabla h| \ll 1$, and the deep water assumption for the wave motion is implied. We also assume that the fluid kinematic viscosity $\nu$ is small, $\gamma = \sqrt{\nu k^2/\omega} \ll 1$, where $k = \omega^2/g$ is a characteristic wave number. The viscosity of the fluid results in the fact that the wave motion ceases to be potential in a thin viscous sublayer of thickness $\delta \sim \gamma /k$ near the fluid surface. We represent the wave velocity as a sum of potential and vortical terms, $\bm u = \bm u^{\phi} + \bm u^{\psi}$, where $\bm u^{\phi} = \nabla \phi$ corresponds to the potential term and the vortical term $\bm u^{\psi}$ is parametrically smaller near the surface, $|\bm u^{\psi}| \sim \gamma |\bm u^{\phi}|$, and it is absent in the fluid bulk below the viscous sublayer \cite{lamb1975hydrodynamics}. Note that the wave amplitude can be either larger or smaller than the thickness $\delta$ of the viscous sublayer.

\begin{figure}[t]

\includegraphics[width=0.7\linewidth]{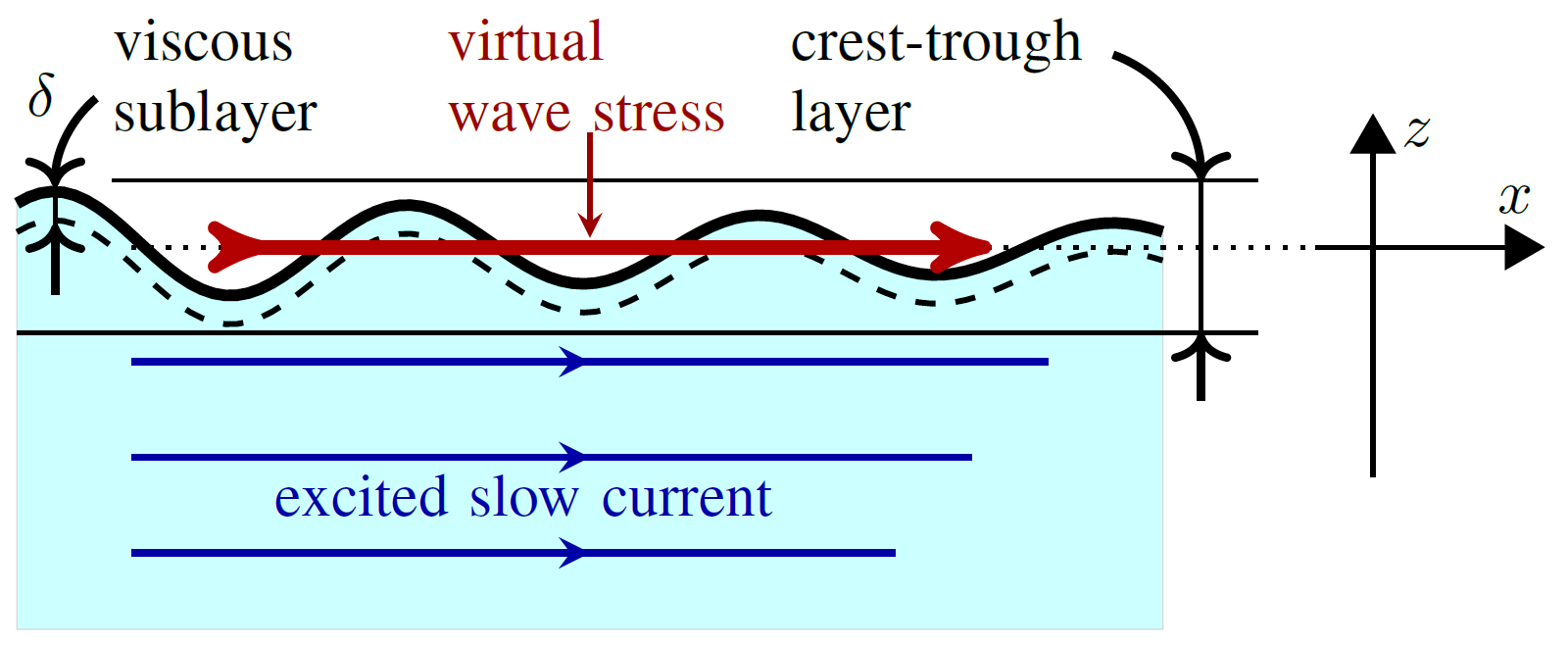}
\caption{Schematic of the slow current generation by a progressive surface wave in a slightly viscous fluid.}
\label{fig:0}
\end{figure}

Next, we denote by $\bm V$ the slow current, which is different from the slow potential contribution to the wave motion (it appears as a result of hydrodynamic nonlinearity and its amplitude is proportional to $\Delta \omega$, see Ref.~\cite{longuet1963effect}), and so the velocity of fluid is $\bm v = \bm u + \bm V$. The characteristic time scale $T$ of the slow current $\bm V$ is much larger than the inverse wave frequency, i.e. $\omega T \gg 1$. Concerning the characteristic length scale $L$ of the slow flow $\bm V$, it does not always far exceed the wavelength, they can be of the same order, see, e.g., Ref.~\cite{parfenyev2019formation}. Also, we assume that the fluid surface remains approximately flat if only the slow current is excited. This means that the Froude number $\mathrm{Fr}=V^2/(gL)$ for the slow current is small.

Our goal is to establish the influence of the wave motion on the slow current $\bm V$. There are two fundamentally different ways for this effect. The first way is due to the fluid viscosity and therefore it is forbidden in an ideal fluid. The wave motion attenuates and its momentum is transferred to Eulerian slow current by the action of the virtual wave stress \cite{longuet1969nonlinear, weber2001virtual}. This stress $\bm \tau$ slowly changes in time, of the second-order in the wave amplitude $h$ and linear in the fluid viscosity $\nu$. Its amplitude can be estimated as the rate of decay of the surface density of momentum $|\bm \pi| = \rho \omega \langle h^2 \rangle$ in progressive wave (which is equal to the Stokes drift integrated over the fluid depth and multiplied by the fluid density $\rho$, see Ref.~\cite{longuet1969nonlinear}), i.e. $|\bm \tau| \sim \nu k^2 |\bm \pi|$. Here and below angle brackets $\langle \cdots \rangle$ mean the extraction of slow harmonics with frequencies of the order of or less than $\Delta \omega$. The virtual wave stress $\bm \tau$ is applied near the fluid surface and therefore its action produces the slow current with non-zero vorticity, see Fig.~\ref{fig:0}. One can say that this vorticity is created in the viscous sublayer due to the fluid viscosity and hydrodynamic nonlinearity, and then it spreads downward in the fluid bulk through the viscous diffusion \cite{longuet1953mass,parfenyev2019formation}. The concept of the virtual wave stress is inevitably associated with the fluid viscosity, since the generation of vorticity in an ideal fluid by the potential wave motion is forbidden due to Kelvin's theorem \cite{falkovich2011fluid}. In this paper, we focus on finding the virtual wave stress $\bm \tau$ in the case of excitation of almost monochromatic waves propagating in arbitrary directions.

The second way of the influence is amplification of the vorticity associated with the already excited slow current $\bm V$ due to its interaction with the wave motion (see, e.g., paper~\cite{ardhuin2017comments} and references therein, as well as recent experimental works \cite{savelyev2012turbulence,d2014quantifying}). The fluid viscosity is not important here. The interaction between the slow current and the wave motion is localized at the wave penetration depth $\sim 1/k$, where the wave motion is potential. The effect is associated with the wave scattering on the inhomogeneities of the slow vortical flow $\bm V$ due to hydrodynamic nonlinearities \cite{phillips1959scattering}. In this paper, we assume that the effect is negligible as compared to the action of the virtual wave stress. The scattering in the fluid bulk is negligible if the condition $V/L \ll \nu k^2$ is satisfied, see Sec.~\ref{sec:WCI} below, and the scattering on a curved surface can be neglected under the additional condition ${\mathrm{Fr}}\ll \gamma^2$, see the explanation after equation (\ref{corr}) in Sec.~\ref{sec:WVS}.

Theoretical analysis of the problem is based on the use of integral forms of the continuity equation and the Navier-Stokes equation, which are respectively the laws of mass and momentum conservation. We introduce the momentum flux density tensor,
\begin{equation}\label{eq:Pi}
\Pi_{ij} = p \delta_{ij} + \rho v_i v_j - \rho \nu (\partial_i v_j + \partial_j v_i),
\end{equation}
which is the $i$-th component of the amount of momentum flowing in unit time through unit area perpendicular to the $j$-axis \cite{landau1987course}. Here $p$ is the pressure, $\delta_{ij}$ is the Kronecker delta and the fluid is assumed to be incompressible, $\mathrm{div}\, \bm v = 0$. Now, we can write the mass conservation law
\begin{equation}\label{eq:incompress}
  \partial_t[\theta(h-z) \rho] + \partial_j [\theta(h-z) \rho v_j] = 0,
\end{equation}
and the Navier-Stokes equation
\begin{equation}\label{eq:Navier-Stokes}
  \partial_t[\theta(h-z) \rho v_i] + \partial_j [\theta(h-z) \Pi_{ij}] = -\delta_{iz} \theta(h-z) \rho g,
\end{equation}
where $\theta(h-z)$ is the Heaviside step function and we sum over the repeated Latin indices that run through the values $\{x,y,z\}$. Note that both these equations are applicable in the whole space and contain exact boundary conditions, which should be found from the requirement that the coefficients before the Dirac delta function $\delta(h-z) = \theta'(h-z)$ are equal to zero, see Appendix~\ref{sec:A} for detail.

In the fluid bulk, below the viscous sublayer, the wave motion is potential, and since we have neglected its scattering on the inhomogeneities of the slow current, the wave motion cannot change the vorticity of the flow. It means that the slow vortical flow $\bm V$ is described by the usual Navier-Stokes equation in the fluid bulk
\begin{equation}\label{eq:Navier-Stokes-V}
  \partial_t \bm V + (\bm V \nabla) \bm V + \nabla P/\rho - \nu \nabla^2 \bm V = 0,
\end{equation}
supplemented by the incompressibility condition $\mathrm{div}\,\bm V = 0$, which follows from equation (\ref{eq:incompress}). Here $P$ is contribution to the pressure, associated with the slow current $\bm V$. Note that due to the incompressibility condition $\nabla^2 P = - \rho (\partial_i V_j) (\partial_j V_i)$.

The Navier-Stokes equation (\ref{eq:Navier-Stokes-V}) for the slow current $\bm V$ should also be supplemented by the boundary conditions. Since we have assumed that the fluid surface remains flat if only the slow current is excited, these conditions can be posed at fixed virtual boundary $z = 0$ corresponding to the unperturbed fluid surface. As was explained earlier, due to the wave motion and the fluid viscosity, the virtual wave stress $\bm \tau$ is applied to this boundary and so it is not free. The presence of full divergency in equation (\ref{eq:Navier-Stokes}) allows one to exploit the momentum conservation law in integral form for crest-trough layer which is only partially filled with the fluid, see Fig.~\ref{fig:0}, and obtain the explicit expression for the virtual wave stress in terms of the excited wave motion. Details of the calculation are presented in Sec.~\ref{sec:WVS}.

\section{Wave-Current Interaction in the Fluid Bulk}\label{sec:WCI}

Before proceeding to the calculation of the virtual wave stress, we discuss the condition when the wave-current interaction can be neglected. The condition is equivalent to the requirement that the scattering of wave motion by the slow flow inhomogeneities due to the hydrodynamic nonlinearity is small as compared with the viscous wave damping. Here we consider the influence of nonlinearity in the fluid volume. If the scattering is weak, the wave motion is potential below the viscous sublayer and the velocity of fluid is equal to $\bm v^0 = \bm u^{\phi} + \bm V$. In general case, the Navier-Stokes equation (\ref{eq:Navier-Stokes}) in the fluid bulk has the form
\begin{equation}\label{eq:Navier-Stokes-uV}
  \rho \partial_t v_i = -\partial_j (\Pi_{ij}^0 + \delta \Pi_{ij}) - \delta_{iz} \rho g,
\end{equation}
where the momentum flux $\Pi_{ij}^0$ corresponds to the velocity field $\bm v^0$ and $\delta \Pi_{ij} = \Pi_{ij} - \Pi_{ij}^0$. First, we consider the terms in equation (\ref{eq:Navier-Stokes-uV}) that have a characteristic frequency of $\omega$. The scattering of the wave motion on the inhomogeneities of the slow current $\bm V$ corresponds to the term $\rho u_j^{\phi} \partial_j V_i$ in $\partial_j \Pi_{ij}^0$, which leads to the deviation of the wave flow $\bm u$ from the potential flow $\bm u^{\phi}$ at depth of the order of $1/k$. This difference can be estimated as $\bm u^{scat} \sim h(V/L)$, and let us stress that the vortical correction $\bm u^{scat}$ is localized at the depth of $1/k$ and has nothing common with the vortical correction $\bm u^{\psi}$ localized in the viscous boundary sublayer near the fluid surface.

Next, the vortical correction $\bm u^{scat}$ produces contribution in the average value of the momentum flux difference $\langle \delta \Pi_{ij} \rangle$, which can be estimated as $\rho \langle u^{\phi} u^{scat} \rangle \sim \rho \omega \langle h^2 \rangle (V/L)$. The influence of this additional term on the flow can be neglected, if it is less than the virtual wave stress $|\bm \tau| \sim \rho \nu \omega k^2 \langle h^2 \rangle$. Thus, we obtain the condition for the slow current gradient $V/L \ll \nu k^2$, which is assumed to be fulfilled in this paper. The condition is equivalent to the requirement that the wave scattering length on slow current inhomogeneities is greater than the propagation length of the wave $l_\nu \sim \omega/(\nu k^3)$, which is determined by viscous damping. Indeed, one has $|\nabla {\bm V}| l_\nu \ll c_g$, where $c_g=\omega/(2k)$ is the group velocity of the waves, that is the variation of velocity ${\bm V}$ is negligible for a propagating wave.

\section{Virtual Wave Stress}\label{sec:WVS}

\begin{figure}[t]

\includegraphics[width=0.7\linewidth]{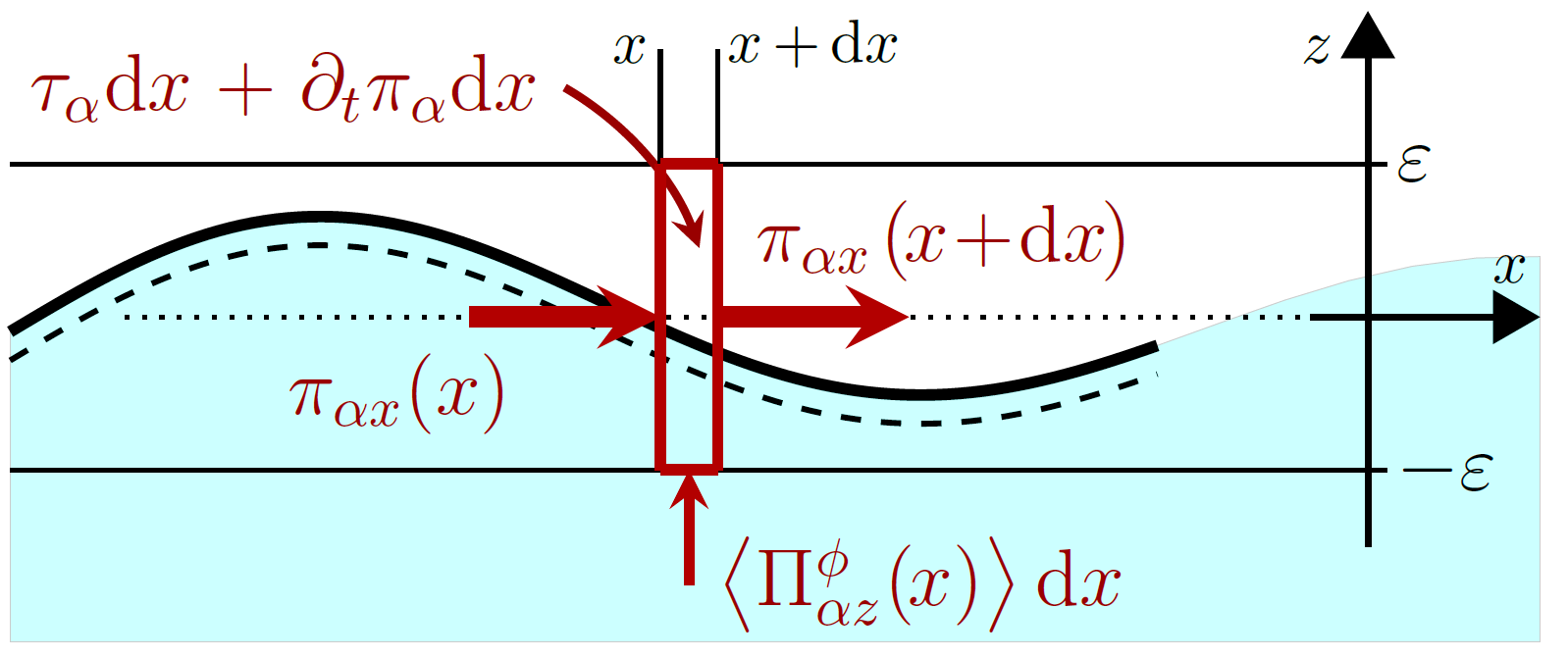}
\caption{A change in the horizontal component of the wave motion momentum generates a surface force $\tau_{\alpha}$ (virtual wave stress) that excites a slow flow.}
\label{fig:1}
\end{figure}

The bulk equation (\ref{eq:Navier-Stokes-V}) has to be supplemented by three boundary conditions for the velocity field $\bm V$ posed at a fixed virtual boundary $z = 0$ corresponding to the unperturbed fluid surface. The vertical velocity can be estimated as divergence of Stokes mass transport for progressive wave, $V_z\vert_{z=0} \sim \nu k^3 \langle h^2 \rangle$, and then in the leading order the first boundary condition is $V_z\vert_{z=0} \approx 0$ (see also Sec.~\ref{sec:Ex} for examples and Sec.~\ref{sec:discussion} for a more thorough analysis). To obtain other boundary conditions, we introduce the Heaviside step function for a fixed virtual boundary $\theta^0 \equiv \theta(-z)$ and for the real boundary $\theta \equiv \theta(h-z)$, and consider the horizontal component of the Navier-Stokes equation (\ref{eq:Navier-Stokes}), which can be rewritten as
\begin{equation}\label{eq:VWS_eq}
   \partial_t(\theta^0 \rho v_{\alpha}^0) + \partial_j ( \theta^0 \Pi_{\alpha j}^0 ) = -\rho \partial_t (\delta v_{\alpha}) - \partial_j (\delta \Pi_{\alpha j}).
\end{equation}
Here and below Latin indices take the values $\{x,y,z\}$ and Greek indices take only $\{x,y\}$, we sum over the repeated indices, $\delta {\bm v} = \theta {\bm v}-\theta^0{\bm v}^0$, $\delta \Pi_{ij} = \theta \Pi_{ij} - \theta^0\Pi^0_{ij}$ and the momentum flux $\Pi_{ij}^{0}$ corresponds to velocity field ${\bm v}^0 = \bm u^{\phi} + \bm V$. The left-hand side of this equation corresponds to the Navier-Stokes equation with the fixed flat virtual boundary and a purely potential flow in surface waves. The virtual boundary partially extends beyond the fluid, therefore we do an analytic continuation of the velocity ${\bm v}^0={\bm V}+{\bm u}^\phi$ toward the boundary. Since the boundary $z=0$ is virtual and not real, the left-hand side of equation (\ref{eq:VWS_eq}) is not zero. It is equal to $-\delta(z) \Pi_{\alpha z}^0$ and this imbalance is compensated by the right-hand side, which is non-zero only near the fluid surface, inside the region $|z|<\varepsilon$, see Fig.~\ref{fig:1}. The constant $\varepsilon$ is much less than the wavelength, $k \varepsilon \ll 1$, but the plane $z=-\varepsilon$ is always below the fluid surface and the vortical part of velocity associated with waves $\bm u^{\psi}$ is always negligible at the depth $z=-\varepsilon$. Such a constant exists because we assumed that $kh \ll 1$ and $k \delta \ll 1$.

Next, we average equation (\ref{eq:VWS_eq}) over the wave oscillations and approximate the right-hand side as $\delta(z) F_\alpha$. The effective boundary conditions imposed on the virtual boundary should be obtained by equating the overall coefficient before $\delta(z)$ in relation (\ref{eq:VWS_eq}) to zero, i.e. $\langle\Pi^0_{\alpha z}\rangle\big\vert_{z=0}=-F_\alpha$. To simplify calculations, we choose an inertial reference frame in which the horizontal components of the slow current are equal to zero near the surface at a given position and time, $V_\alpha\vert_{z=0}=0$. In this reference frame, not only the gradient of the slow current is small, but also its absolute value, and therefore its interaction with the wave motion and with itself can be neglected. Below, in Sec.~\ref{sec:discussion}, we will discuss how to take into account additional terms that correspond to advection by a constant horizontal velocity associated with a moving reference frame, and thereby restore the Euler invariance inherent in the original equations. Note also that the considered case corresponds to the initial stage of the slow current generation by surface waves, if it was initially absent.

To find $F_\alpha$, one needs to integrate the right-hand side of time-averaged equation (\ref{eq:VWS_eq}) in $Z$-direction, after which we obtain
\begin{equation}\label{BC-project}
\langle \Pi_{\alpha z}^V + \Pi^\phi_{\alpha z}\rangle \big\vert_{z=0} = \partial_\beta \int\limits_{-\infty}^{+\infty}\mathrm{d}z \left\langle\delta \Pi_{\alpha \beta}\right\rangle + \partial_t\int\limits_{-\infty}^{+\infty}\mathrm{d}z\langle\rho\,\delta v_\alpha\rangle,
\end{equation}
where $\Pi_{\alpha z}^0 = \Pi_{\alpha z}^V + \Pi^\phi_{\alpha z}$, and $\Pi^V_{\alpha z}$ and $\Pi^\phi_{\alpha z}$ are momentum fluxes corresponding to velocity fields ${\bm V}$ and ${\bm u}^\phi$ respectively. Note that during subsequent calculations, we should keep only linear contributions in viscosity, which are leading in the parameter $\gamma \ll 1$, and should limit our analysis to second-order in wave amplitude, following the mentioned estimates for the virtual wave stress. All higher-order corrections should be neglected. We can also skip all linear in the wave amplitude contributions, because the wave velocity field satisfies the linear equations and these terms will cancel each other. Also note, that we may not think about terms at a double frequency $2 \omega$ produced due to hydrodynamic nonlinearity, since they are separated from all equations and boundary conditions and form a closed subsystem of equations. Accordingly, we can put $\bm u^{\phi} = \nabla (\phi^{(1)}+\phi^{(2)})$, where $\phi^{(1)}$ is the wave potential in the linear approximation and $\phi^{(2)}$ is the slow second-order correction proportional to $\Delta \omega$, see Ref.~\cite{longuet1963effect}. Some useful well-known expressions for the wave motion are summarized in Appendix~\ref{sec:B}.

Now we proceed to the analysis of the left-hand side of equation (\ref{BC-project}). In our reference frame $V_\alpha\vert_{z=0}=0$, and therefore we obtain $\Pi^V_{\alpha z}\vert_{z=0} = -\rho \nu \partial_z V_\alpha\vert_{z=0}$. We have neglected $\rho \nu\partial_\alpha V_z\vert_{z=0}$ because $V_z\vert_{z=0} \sim \nu k^3 \langle h^2 \rangle$ and the term contains an additional smallness in parameter $\gamma^2 \ll 1$. Next, the mean value of the momentum flux associated with the potential wave motion is equal to
\begin{eqnarray}\label{Pi_through_phi}
    \left\langle
    \Pi^{\phi}_{\alpha z}
    \right\rangle\big\vert_{z=0}
    =
    \rho
    \left\langle
    \partial_\alpha\phi^{\scriptscriptstyle (1)} \,\partial_z\phi^{\scriptscriptstyle (1)}
    \right\rangle\big\vert_{z=0}
    -2\rho\nu \partial_{\alpha z}\langle \phi^{\scriptscriptstyle (2)}\rangle \big\vert_{z=0}.
\end{eqnarray}
Since the last term contains an explicit factor $\nu$, the velocity potential $\langle \phi^{\scriptscriptstyle (2)}\rangle$ should be found in the limit of an ideal fluid without the viscous correction. According to Ref.~\cite{longuet1963effect}, we can estimate $\partial_{\alpha z}\langle \phi^{\scriptscriptstyle (2)}\rangle \sim \Delta \omega k^2 \langle h \rangle^2$ and therefore $\rho\nu\partial_{\alpha z}\langle \phi^{\scriptscriptstyle (2)}\rangle/|{\boldsymbol \tau}| \sim \Delta\omega/\omega \ll 1$ and the term should be neglected. Note also, that the potential $\phi^{\scriptscriptstyle (2)}$ satisfies Laplace equation and boundary conditions $\nabla \phi^{\scriptscriptstyle (2)} \to 0$ when $z\to-\infty$ and $(\partial_t^2+g\hat k)\phi^{\scriptscriptstyle (2)}= -\partial_t (\nabla \phi^{\scriptscriptstyle (1)})^2$ at $z=0$, where $\hat k = \sqrt{-\partial_{\alpha} \partial_{\alpha}}$ is the wave number operator. The presence of the full-time derivative in the last boundary condition leads to the fact that $\partial_{\alpha z}\langle \phi^{\scriptscriptstyle (2)}\rangle \vert_{z=0}$ itself is the full-time derivative of limited in time quantity. This means that the time integral of this quantity is also limited. Thus, the last term in equation (\ref{Pi_through_phi}) cannot lead to the excitation of a significant slow current at long times. Finally, let us verify that the cross-contribution to the momentum flux $\Pi^{V\!\text{-}\phi}_{\alpha z}$ can be neglected. The contribution is $\rho V_z\langle u^{\scriptscriptstyle (2)}_\alpha\rangle\vert_{z=0} \sim \rho\nu \Delta\omega k^4 \langle h^2\rangle^2$ and it is of the fourth-order in wave amplitude.

Next, we turn out to the analysis of the right-hand side of equation (\ref{BC-project}). The last term without $\partial_t$ is the horizontal momentum surface density $\pi_\alpha$ for the wave motion
\begin{equation}\label{pi_a}
    \pi_\alpha
    =
    \rho\left\langle\int\limits^{h}_{0} \mathrm{d}z\,u^\phi_\alpha + \int\limits^{h}_{-\infty} \mathrm{d}z\,u^\psi_\alpha\right\rangle
    =
    \rho\langle h \,\partial_\alpha \phi^{\scriptscriptstyle (1)}\rangle \big\vert_{z=0}.
\end{equation}
The vortical part of the velocity $u^\psi_\alpha$ produces $\rho \langle \psi_\alpha^{\scriptscriptstyle (1)} \vert_{z=h}\rangle$, which should be neglected due to time averaging, see equation (\ref{psi}). Note that in our analysis, the second-order slow vortex contribution $\psi_\alpha^{\scriptscriptstyle (2)}$ generated by waves due to hydrodynamic nonlinearity is included in the definition of the slow flow $\bm V$, so there are no additional terms in the equation. In the case of plane wave in an ideal fluid, $\pi_{\alpha}$ corresponds to the mass-transport through the wave total cross-section, since the corresponding integral is accumulated at the fluid surface in the Euler description, see Ref.~\cite{longuet1969nonlinear}. The time derivative takes into account the possible time decay of this quantity due to the fluid viscosity.

The first term in the right-hand side of equation (\ref{BC-project}) without $\partial_\beta$ is surface density of the horizontal momentum flux $\pi_{\alpha\beta}$, and the spatial derivative takes into account the possible spatial decay of this quantity due to the fluid viscosity, see also Ref.~\cite{weber2001virtual},
\begin{equation}\label{pi_alpha-beta}
    \pi_{\alpha\beta}
      =
    \left\langle \int_0^{h}
    \big(p_u^{\scriptscriptstyle (1)}\delta_{\alpha\beta}
    -
    \rho \nu (\partial_\alpha u^\phi_\beta + \partial_\beta u^\phi_\alpha)\big)
    \mathrm{d}z\right\rangle
    +
    \left\langle
    \int_{-\infty}^{h}
    \Big(\rho(u^\psi_\alpha u^\phi_\beta + u^\psi_\beta u^\phi_\alpha)
    +
    p_u^\psi \delta_{\alpha\beta}\Big)
    \mathrm{d}z
    \right\rangle.
\end{equation}
Here, we neglected the second-order contributions in the wave amplitude that contain the small factor $\gamma^3$ and denoted by $p_u^\psi $ the pressure associated with the vortex flow $\bm u^\psi$, which is non-zero only inside the viscous sublayer. As it turns out, the second term in equation (\ref{pi_alpha-beta}) should also be neglected, and the first term gives the result
\begin{equation}\label{pi_ab_found}
    \pi_{\alpha\beta}
    =
    -\rho\left\langle h\,\partial_t \phi^{\scriptscriptstyle (1)}\vert_{z=0} + gh^2/2\right\rangle\delta_{\alpha\beta}
    -
    2\rho \nu \left\langle  h\, \partial_{\alpha\beta} \phi^{\scriptscriptstyle (1)}\right\rangle\vert_{z=0},
\end{equation}
where we have used Bernoulli equation (\ref{Bernoulli}) in the linear approximation, which is valid for the viscous fluid as well, see Ref.~\cite[\S 349]{lamb1975hydrodynamics}.

To justify the neglect of the second term in equation (\ref{pi_alpha-beta}), we first consider the term containing the pressure $p_u^\psi$. The contribution satisfies $\nabla^2 p_u^\psi = -\rho [\partial_i u_k \partial_k u_i - (\partial_{ik}\phi)^2]$, and the right-hand side of this equation can be estimated as $\rho (\omega k h)^2 $ and it is localized in the viscous sublayer. Below it, the wave motion is potential and satisfies the Euler equation, which can be integrated, leading to the Bernoulli equation (\ref{Bernoulli}) in the fluid bulk. Thus, the correction to pressure $p_u^{\psi}$ is non-zero only inside the thin viscous sublayer and it can be estimated as $\langle p_u^\psi\rangle \sim \gamma^2 \rho \omega^2 \langle h \rangle^2$. The integration across the viscous sublayer in relation (\ref{pi_alpha-beta}) produces one more factor $\gamma$, which makes the contribution negligible. Second, let us analyze the remaining contribution to the second term of equation (\ref{pi_alpha-beta}), which is equal to $\delta\pi_{\alpha\beta} = \rho\int^h\mathrm{d}z  \langle\partial_{z}\psi_\alpha^{\scriptscriptstyle (1)}\,\partial_\beta \phi^{\scriptscriptstyle (1)} + \partial_{z}\psi_\beta^{\scriptscriptstyle (1)}\,\partial_\alpha\phi^{\scriptscriptstyle (1)}\rangle = \rho \langle \psi_\alpha^{\scriptscriptstyle (1)} \, \partial_\beta\phi^{\scriptscriptstyle (1)} + \psi_\beta^{\scriptscriptstyle (1)} \, \partial_\alpha\phi^{\scriptscriptstyle (1)}\rangle\big\vert_{z=h}$. It follows from equation (\ref{psi}) and the wave equation (\ref{wave_equation}) that $\psi_\alpha^{\scriptscriptstyle (1)}\vert_{z=h} = (2\nu/g)\partial_{t\alpha}\phi^{\scriptscriptstyle (1)}\vert_{z=0}$ in the limit of small viscosity, and hence the contribution is the full-time derivative, $\delta\pi_{\alpha\beta}= (2\rho\nu/g) \partial_t \langle \partial_\alpha\phi^{\scriptscriptstyle (1)} \,\partial_\beta \phi^{\scriptscriptstyle (1)} \rangle\vert_{z=0} \sim \rho \nu \, \Delta\omega\,k\langle h^2\rangle $. This expression is small as $\Delta \omega/\omega \ll 1$ compared to the last term in equation (\ref{pi_ab_found}), and it should be neglected as it was done for the last term in relation (\ref{Pi_through_phi}).

Now we consider the remaining terms in the right-hand side of equation (\ref{BC-project}) that have not yet been discussed and show that they can also be neglected. These terms are the result of a separate averaging of either the upper integration limit $h$ or the integrand. Because the corrections arising from the vortical part $\bm u^{\psi}$ of wave flow inside the viscous sublayer have already been taken into account and neglected, we can replace $ \Pi_{ij} $ with $\Pi^0_{ij}$ in the right-hand side of equation (\ref{BC-project}) and then the integrand in it becomes equal to
\begin{eqnarray}\nonumber
    (\theta-\theta^0)(\rho \partial_t v^0_\alpha + \partial_\beta\Pi^0_{\alpha \beta} )
    +
    \delta(z-h) (\Pi_{\alpha\beta}^0 \partial_\beta h + \rho v^0_\alpha \partial_t h)
    =
    \\[5pt]\label{corr0}
    =
    -(\theta-\theta^0)\partial_z\Pi^0_{\alpha z}
    +
    \delta(z-h) (\Pi_{\alpha\beta}^0 \partial_\beta h + \rho v^0_\alpha \partial_t h).
    \hskip20pt
\end{eqnarray}
The equality in equation (\ref{corr0}) is valid since nonlinear interaction between waves and slow current can be neglected in the fluid bulk, and we remind that under the assumption $\langle \Pi^0_{\alpha z}\rangle = \Pi^V_{\alpha z} + \langle \Pi^\phi_{\alpha z}\rangle$. Since we have already taken into account the correlations between the upper limit of integration $h$ and integrable in the right-hand side of equation (\ref{BC-project}), the remaining terms are
\begin{eqnarray}\nonumber
    &-\langle h\rangle \partial_z \Pi^V_{\alpha z}\vert_{z=0}
    -
    h^V \partial_z\langle \Pi^\phi_{\alpha z}\rangle\vert_{z=0}
    +
    \langle \Pi^\phi_{\alpha\beta}\vert_{z=h}\rangle \partial_\beta h^V+&
    \\[5pt]\label{corr}
    &+ \Pi^V_{\alpha\beta}\vert_{z=h^V}\partial_\beta \langle h^{(2)}\rangle
    + \rho \langle u_\alpha^\phi\vert_{z=h} \rangle \partial_t h^V.&
\end{eqnarray}
Here $\langle h \rangle \approx h^V + \langle h^{\scriptscriptstyle (2)} \rangle$, where $h^V$ is the surface elevation corresponding to the slow current in the absence of waves and $\langle h^{\scriptscriptstyle (2)} \rangle$ is produced by the slow second-order wave motion $\phi^{\scriptscriptstyle (2)}$. We kept only the part $h^V$ from $\langle h\rangle$ in several terms in the right-hand side of relation (\ref{corr}) to omit all contributions proportional to $h^4$. The first term can be estimated as $(\langle h\rangle/L)\Pi^V_{\alpha z}$, and it is small as compared to the term $\Pi^V_{\alpha z}$, which is already taken into account in the left-hand side of equation (\ref{BC-project}). The third term in relation (\ref{corr}) is small as compared to the virtual wave stress if the condition $\mathrm{Fr}\ll \gamma^2$ is fulfilled. Physically this means that the wave scattering on the curved surface is negligible. Under the condition, $h^V$ can be considered constant in the second term, so the sum of this term with the second term in the left-hand side of equation (\ref{BC-project}) means that $h^V$ simply changes the zero elevation point for fluid when describing the surface wave motion. Considering the fourth term, one can estimate $\Pi^V_{\alpha\beta}\sim \rho \nu V/L$, including the contribution from pressure which can be estimated from the boundary condition $P = 2 \rho \nu \partial_z V_z$. Therefore, the term can be neglected as compared to the left-hand side of equation (\ref{BC-project}) due to $\partial_\beta \langle h^{(2)}\rangle \ll 1$. The last term has a similar nature to the second one. It takes into account that the fluid surface can move upwards as a whole with the local velocity $\partial_t h^V$, and it can be eliminated if we choose the corresponding reference frame.

Finally, collecting all the contributions together, we can write equation (\ref{BC-project}) in the form
\begin{equation}\label{BC}
    \rho \nu \partial_z V_\alpha\vert_{z=0}
    =
    -\partial_\beta\pi_{\alpha\beta}
    +\langle \Pi^\phi_{\alpha z}\rangle\vert_{z=h^V}
    -\partial_t \pi_\alpha
    \equiv
    \tau_\alpha,
\end{equation}
which is illustrated in Fig.~\ref{fig:1}. It is convenient to express the final result in terms of the wave elevation, and by substituting relations (\ref{Pi_through_phi},\ref{pi_a},\ref{pi_ab_found}) in equation (\ref{BC}) and using wave equation (\ref{wave_equation}) and relation (\ref{flow_linear}), we obtain
\begin{equation}\label{eq:tau}
    \tau_\alpha = -2 \rho \nu \left\langle k^{-1} (\partial_\beta \partial_t h)  (\partial_{\alpha} \partial_{\beta} h)  + k (\partial_{\alpha} h)  (\partial_t h) \right\rangle.
\end{equation}
As we originally expected, the expression for the virtual wave stress is proportional to the viscosity and the square of the amplitude of the wave motion. Note that the term $-\partial_t\pi_\alpha$ in equation (\ref{BC}) is the full-time derivative. And although its amplitude is not small as compared to $|{\boldsymbol \tau}|$ in the general case, it can be neglected for long times, since it cannot generate a significant slow current. However, taking this term into account, we can clearly show that the virtual wave stress (\ref{eq:tau}) is proportional to the viscosity.

To summarize, the bulk equation (\ref{eq:Navier-Stokes-V}) has to be supplemented by the boundary conditions
\begin{equation}\label{eq:BCBC}
  \rho \nu \partial_z V_{\alpha} \vert_{z=0} = \tau_{\alpha}, \quad V_z\vert_{z=0} = 0,
\end{equation}
posed at the unpertubed fluid surface $z=0$, where the virtual wave stress $\tau_{\alpha}$ is defined by equation (\ref{eq:tau}). These relations imply the fulfillment of two inequalities: $V/L \ll \nu k^2$ and $V^2/(gL) \ll \gamma^2$, which mean the weakness of the slow flow. Note that the divergence of the virtual wave stress is zero, $\partial_\alpha \tau_\alpha = 0$, and the curl of both sides of equation (\ref{eq:BCBC}) gives the boundary condition for $Z$-derivative of the vertical vorticity (the part which spreads outside the viscous sublayer) that was previously obtained in Ref.~\cite[Eq.~(12)]{filatov2016nonlinear}.

\section{Examples}\label{sec:Ex}

To illustrate our result let us find the slow flow generated by a propagating wave \cite{longuet1953mass} and two orthogonal standing waves \cite{parfenyev2019formation}. Following these references, we assume that the wave motion is stationary in time due to external pumping and the resulting slow current $\bm V$ is rather weak, so the nonlinear term in the Navier-Stokes equation (\ref{eq:Navier-Stokes-V}) can be neglected. In the first case, the wave elevation is
\begin{equation}\label{eq:h1}
  h(t,x) = H \cos(kx-\omega t),
\end{equation}
where $H=H_0 \exp(-4 \gamma^2 kx)$ decays in space due to the fluid viscosity. By using equation (\ref{eq:tau}), we find that the surface force density exciting the slow current is equal to $\tau_x = 2 \rho \nu \omega (kH)^2$. Therefore, we obtain that at the virtual boundary
\begin{equation}\label{bc1}
  \partial_z V_x\vert_{z=0} = 2 \omega (kH)^2, \quad V_z\vert_{z=0} = 0.
\end{equation}
Next, the slow flow penetrates the fluid bulk due to viscous diffusion and ultimately reaches the bottom of the system, which is located at $z=-d$. To find the stationary solution we need to impose here the usual no-slip boundary condition
\begin{equation}\label{bc2}
  V_x\vert_{z=-d}=0, \quad V_z\vert_{z=-d}=0.
\end{equation}
Note that $d \gg 1/k$ to satisfy the deep-water approximation for surface waves. Nevertheless, the slow current reaches the bottom in the stationary regime and therefore its position plays an important role.

To simplify the problem, we suppose that the fluid depth $d$ is much less than the wave propagation length $l_{\nu} = 1/(4 \gamma^2 k)$. In this case, in the main approximation with respect to parameter $d/l_{\nu} \ll 1$, one can assume that the virtual wave stress $\tau_x$ acting on the fluid surface is constant. This means that the considered system is homogeneous along the direction of wave propagation and therefore nothing depends on the $x$-coordinate. Following Ref.~\cite{longuet1953mass}, we also assume that total horizontal transport is zero,
\begin{equation}\label{eq:int}
  \int_{-d}^{0} dz \, V_x(z) = 0.
\end{equation}
As one can see later, the correction to this expression associated with the Stokes drift can be neglected since $d \gg 1/k$.

Now, let us proceed to calculations. The incompressibility condition leads to $\partial_z V_z = 0$ and due to the condition $V_z(-d)=0$ this means that $V_z = 0$. Next, from $z$-component of equation (\ref{eq:Navier-Stokes-V}) we obtain $\partial_z P = 0$, which means that pressure $P$ does not depend on $z$-coordinate. Considering $x$-component of equation (\ref{eq:Navier-Stokes-V}) and using boundary conditions (\ref{bc1}) and (\ref{bc2}) together with expression (\ref{eq:int}), we find the stationary solution
\begin{equation}\label{eq:result}
  V_x(z) = \dfrac{\omega (kH)^2 d}{2}  \left( \frac{3z^2}{d^2} + \frac{4z}{d} +1 \right).
\end{equation}
The presented derivation leads to the same result as in Ref.~\cite[Eq.(305)]{longuet1953mass}. Note that the applicability condition $V/L \ll \nu k^2$ is equivalent to $H \ll \delta$, i.e. the wave amplitude should be much smaller than the thickness of the viscous sublayer. In the opposite case, the expression (\ref{bc1}) is valid only at the initial stage. In the course of further evolution, the slow flow becomes so strong that it is necessary to take into account its interaction with the waves.

In the second example, we consider the slow current generated by two orthogonal standing surface waves \cite{parfenyev2019formation}. The surface elevation is given by
\begin{equation}\label{eq:h2}
  h = H_1 \cos(\omega t) \cos (kx) + H_2 \cos(\omega t + \vartheta) \cos (ky),
\end{equation}
where $H_1$ and $H_2$ are the amplitudes of the waves, and $\vartheta$ is the phase shift between them. As in the previous case, one does not need to take into account viscous corrections to this expression corresponding to the wave spatial decay, since they will produce parametrically smaller contribution to the generated slow flow.

Following Ref.~\cite{parfenyev2019formation}, we will describe the corresponding slow current in terms of the vertical vorticity, $\Omega_E = \partial_x V_y - \partial_y V_x$. Using relations (\ref{eq:Navier-Stokes-V}) and (\ref{eq:BCBC}) and neglecting the nonlinear term in the Navier-Stokes equation, one finds the bulk equation
\begin{equation}\label{eq:vorticity}
  \partial_t \Omega_E - \nu \nabla^2 \Omega_E = 0,
\end{equation}
which has to be supplemented by the boundary conditions
\begin{equation}\label{eq:bc_crossed}
  \rho \nu \partial_z \Omega_E\vert_{z=0} = \epsilon_{\alpha \beta} \partial_{\alpha} \tau_{\beta}, \quad \Omega_E\vert_{z \rightarrow -\infty}=0,
\end{equation}
where $\epsilon_{\alpha \beta}$ is unit antisymmetric tensor and the Greek indices run over $\{x,y\}$. By substituting the wave elevation (\ref{eq:h2}) to equation (\ref{eq:tau}), we obtain
\begin{equation}
  \epsilon_{\alpha \beta} \partial_{\alpha} \tau_{\beta} = -2 \rho \nu \omega k^3 H_1 H_2 \sin (kx) \sin(ky) \sin \vartheta.
\end{equation}
Therefore, equation (\ref{eq:vorticity}) with boundary conditions (\ref{eq:bc_crossed}) literally coincides with the equations (7) and (8) in Ref.~\cite{parfenyev2019formation} (under the assumption of zero compression modulus of the surface film) and has exactly the same solution. In the stationary regime, one finds
\begin{equation}
  \Omega_E = - \sqrt{2} e^{kz\sqrt{2}} H_1 H_2 \omega k^2 \sin (kx) \sin (ky) \sin \vartheta.
\end{equation}
The applicability condition $\Omega_E \ll \nu k^2$ leads to the same restriction $H \ll \delta$ for the wave amplitude in the stationary regime.

The advantage of our method is its relative simplicity due to the lack of the need to resolve the fine details of the viscous sublayer, since it does not produce any contribution during calculations. Moreover, our approach has a clear physical meaning because it is based directly on the momentum conservation law, which makes the phenomenon of generation of slow current by surface waves similar to, for example, the radiation pressure of light.

\section{Discussion}\label{sec:discussion}

In this section, we would like to discuss some additional issues. The first issue is related to the boundary condition for the vertical velocity $V_z$ of slow current under the assumption that the fluid surface remains approximately flat if only the slow current is excited. The second issue is related to the fact that up to this point we have solved the problem in a special reference frame in which the slow current is small. The initial equations possess Euler invariance, and therefore, our theoretical scheme can be generalized to an arbitrary reference frame moving with some horizontal velocity.

To obtain the boundary condition at the virtual boundary $z=0$ for the vertical velocity $V_z$, we consider the continuity equation (\ref{eq:incompress}) in the framework of the same approach that we used earlier to obtain and analyze the equation (\ref{eq:VWS_eq}). First, we rewrite the continuity equation in the form
\begin{equation}\label{continuity0}
    \partial_t(\theta^0\rho) + \partial_i(\theta^0\rho v^0_i)
    =
    -\partial_t\big((\theta-\theta^0)\rho\big)
    -\partial_i(\rho \,\delta v_i),
\end{equation}
and let us remind that we are still working in the reference frame where $V_\alpha\vert_{z=0}=0$ at a given position and time. Next, we proceed according to the scheme that led to the boundary condition (\ref{eq:tau}). We average equation (\ref{continuity0}) over fast oscillations. Then we approximate the right-hand side of this equation as $\delta(z)Q$ and equate the total coefficient before $\delta(z)$ to zero. As a result, we obtain equation $\rho \langle v_z^0\rangle\vert_{z=0} = -Q$, which leads to
\begin{equation}\label{VzEq}
    V_z\vert_{z=0}
    =
    \left(\partial_t\langle h^{(2)}\rangle
    -
    \langle u_z^{(2)}\rangle \vert_{z=h^V}
    +
    \frac{\partial_\alpha\pi_\alpha}{\rho}\right)
    +
    \partial_t h^V.
\end{equation}
The round bracket is equal to zero in the case of an ideal fluid. One can check this using the results for $ h^{(2)}$ and $u_z^{(2)}=\partial_z \phi^{(2)}$ obtained in Ref.~\cite{longuet1963effect}. However, the bracket becomes non-zero and proportional to the viscosity for real fluid. We assume that $\phi^{(2)}$ is the full-time derivative not only for an ideal but also for a viscous fluid as well. In particular, $\phi^{(2)}$ should vanish in the limit $\Delta\omega\to0$. Then the role of the first and the second terms in the round bracket is analogous to the role of term $\partial_t\pi_\alpha$ in relation (\ref{BC}): these terms compensate full-time derivative which is contained in the third term. We do not need to calculate these terms explicitly since only the last term in the round bracket contains the contribution that is not a full-time derivative, and therefore only this contribution is of interest. We use relations (\ref{wave_equation},\ref{flow_linear}) to calculate the contribution and obtain the boundary condition
\begin{equation}\label{BCz}
    V_z\big\vert_{z=0} -  \partial_t h^V = -2\nu k \left\langle (\partial_\alpha h)^2 + 3 (kh)^2 \right \rangle.
\end{equation}

Our consideration implies the slow change in the space of the current $\bm V$, while the absolute value of $h^V$ does not have to be small. From the wave's point of view, $h^V$ remains flat and it determines the level of the unperturbed surface for the wave motion. In the general case, this level can change over time and its changes $\partial_t h^V$ are not necessarily small. For example, one can imagine a vessel filled with water into which fluid is constantly added, so that the average water level rises. The estimate $V_z \vert_{z=0} \sim \nu k^3 \langle h^2 \rangle$ used earlier in the text implicitly assumed that there was no such movement. Now we show that in the general case $V_z \vert_{z=0} - \partial_t h^V \sim \nu k^3 \langle h^2 \rangle$. Then, the cross-contribution to the momentum flux $\rho V_z\langle u^{\scriptscriptstyle (2)}_\alpha\rangle \vert_{z=0}$ from the left-hand side of equation (\ref{BC-project}) can be combined with the last term in equation (\ref{corr}), and we obtain $\rho \langle u^{\scriptscriptstyle (2)}_\alpha\rangle (V_z - \partial_t h^V)\vert_{z=0}$ that is small and can be neglected. Thus, the aforementioned changes in the average fluid level do not modify expression (\ref{eq:tau}) for the virtual wave stress obtained previously in the reference frame which moves upwards with the local velocity $\partial_t h^V$.

Next, analyzing examples in Sec.~\ref{sec:Ex}, we suggested that $V_z\vert_{z=0} = 0$ in the leading approximation. This is justified because the viscosity does not enter in the boundary condition for the horizontal velocity, see, e.g., expression (\ref{bc1}), and the viscous correction (\ref{BCz}) to the leading approximation will produce the parametrically smaller slow current. Note also that the already obtained result (\ref{eq:result}) does not satisfy the incompressibility condition, if one reminds that the wave amplitude decays in space due to the fluid viscosity, $H=H_0 \exp(-4 \gamma^2 kx)$. In particular, the discussed viscous correction participates in the resolution of this discrepancy, but it is small and beyond the scope of the present paper.

To restore the Euler invariance for boundary conditions (\ref{BC}) and (\ref{BCz}), we now return to the laboratory reference frame, where $V_\alpha \vert_{z=0}$ is not equal to zero. However, due to assumed restriction $V/L\ll\nu k^2$, the horizontal velocity $V_\alpha$ should be considered almost constant in space. The Euler invariance implies that the partial time derivatives should be replaced with the material derivatives
\begin{equation}\label{EulerINV}
    \partial_t \to \partial_t + V_\beta \partial_\beta
\end{equation}
in boundary conditions (\ref{BC}) and (\ref{BCz}). In particular, the replacement (\ref{EulerINV}) should be implemented in wave equations (\ref{wave_equation}) and (\ref{flow_linear}) in order to take into account the Doppler effect~\cite{stewart1974hf}. This entails the corresponding replacement in equation (\ref{eq:tau}).

Let us demonstrate that our theoretical scheme indeed inherits Euler invariance. If horizontal velocity $V_\alpha$ is not equal to zero, then in our calculations we should replace
\begin{eqnarray}\label{pi_alpha_gen}
    &\pi_\alpha
     \to \pi_\alpha + \rho V_\alpha \langle h^{(2)}\rangle \vert_{z=0},&
    \\[5pt]\label{pi_alpha-beta_gen}
    &\pi_{\alpha\beta}
     \to
    \pi_{\alpha\beta} + \left( V_\alpha \pi_\beta + V_\beta \pi_\alpha + \rho V_\alpha V_\beta \langle h^{(2)}\rangle \right)\vert_{z=0},&
    \\[5pt] \label{Pi_alpha-z_gen}
    &\langle \Pi_{\alpha z}^0\rangle
     \to
    -\rho \nu \partial_z V_\alpha\vert_{z=0} + \langle \Pi^\phi_{\alpha z}\rangle\vert_{z=0} + \rho V_\alpha \big(V_z + \langle u^{(2)}_z\rangle\big)\vert_{z=0},&
\end{eqnarray}
and here we have assumed that $h^V =0$ and $\partial_t h^V=0$. The first rule (\ref{pi_alpha_gen}) results in a change (\ref{EulerINV}) in expression (\ref{VzEq}). Rules (\ref{pi_alpha_gen},\ref{pi_alpha-beta_gen},\ref{Pi_alpha-z_gen}) together with relation (\ref{VzEq}) results in a change (\ref{EulerINV}) in equation (\ref{BC}), as it was expected.

\section{Conclusion}

The attenuation of surface waves due to the fluid viscosity inevitably leads to the excitation of a slow flow $\bm V$. If it turns out to be weak, $V/L \ll \nu k^2$ and $V^2/(gL) \ll \nu k^2/\omega$, for example, at the initial stage of evolution, then one can neglect the scattering of waves by the slow flow inhomogeneities. In this case, the influence of waves on the slow flow is reduced to the action of a virtual wave stress applied at the fluid surface. As a result, the horizontal momentum associated with the wave motion is transferred from the waves to the slow current.

Here, based on the momentum conservation law, we found the explicit expression (\ref{eq:tau}) for the virtual wave stress in terms of the excited wave motion in the deep-water limit. To demonstrate the validity of our approach, we analyzed the slow currents generated by a propagating wave and two orthogonal standing waves. These cases we extensively studied earlier, see Refs.~\cite{longuet1953mass, parfenyev2019formation}, and we were able to reproduce the previously known results, see Sec.~\ref{sec:Ex}.

The main result of this work is the generalization of the expression for the virtual wave stress to the case of excitation of arbitrary wave motion having a narrow spectrum, $\Delta \omega \ll \omega$. It can be used in the numerical simulation to effectively take into account the effect of fast wave motion on a slow flow. Also, our results allow studying the dynamics of slow currents, if the statistics of surface waves is known.

\acknowledgments
This work was supported by the Project 075-15-2019-1893 funded by the Ministry of Science and Higher Education of the Russian Federation. V.M.P. acknowledges support from the Foundation for the Advancement of Theoretical Physics and Mathematics ''BASIS''.

\appendix

\section{General Equations}\label{sec:A}

In this section we demonstrate how to obtain hydrodynamic equations and boundary conditions in the usual form based on relations (\ref{eq:incompress}) and (\ref{eq:Navier-Stokes}). By using $\delta(h-z) = \theta'(h-z)$, one can rewrite equation (\ref{eq:incompress}) in the form:
\begin{equation}
   \rho \delta(h-z)\left[\partial_t h + v_{\alpha} \partial_{\alpha} h - v_z\right] +
   \theta(h-z)\left[\partial_t \rho + \partial_j(\rho v_j)\right] = 0,
\end{equation}
where Latin indices take the values $\{x,y,z\}$, Greek indices take only $\{x,y\}$, and we sum over the repeated indices. The first term corresponds to the kinematic boundary condition posed at the fluid surface, and the second term --- to the mass conservation law inside the fluid. Since we assume $\rho = const$, then from the second term we find the incompressibility condition $\mathrm{div}\, \bm v = 0$.

Similarly, let us consider relation (\ref{eq:Navier-Stokes}). After straightforward calculations, we obtain
\begin{equation}
   \delta(h-z)\left[\rho v_i \partial_t h + \Pi_{i\alpha} \partial_{\alpha} h - \Pi_{iz} \right] +
   \theta(h-z)\left[\rho \partial_t v_i + \partial_j \Pi_{ij} + \delta_{iz} \rho g \right] = 0.
\end{equation}
Here the second term corresponds to the Navier-Stokes equation inside the fluid,
\begin{equation}
  \partial_t \bm v + (\bm v \nabla )\bm v = -\nabla p/\rho + \nu \nabla^2 \bm v + \bm g,
\end{equation}
where we have used expression (\ref{eq:Pi}). To simplify the first term, one needs to utilize the kinematic boundary condition,
\begin{equation}\label{kinematicBC}
  \partial_t h = \left( v_z - v_{\alpha} \partial_{\alpha} h \right)\vert_{z=h},
\end{equation}
which was obtained above, and then we find that $-p \partial_i (z-h) + \sigma_{ij}' \partial_j (z-h) = 0$ at the fluid surface, where $\sigma_{ij}' = \rho \nu (\partial_i v_j + \partial_j v_i)$ is the viscous stress tensor. Introducing the unit vector normal to the fluid surface $\bm n (t,x,y) =  (-\partial_x h, -\partial_y h, 1)/\sqrt{1 + (\nabla h)^2}$, we finally obtain the dynamic boundary condition in the usual form
\begin{equation}\label{dynamicBC}
  (-p n_i + \sigma_{ij}' n_j)\vert_{z=h} = 0.
\end{equation}

\section{Linear Waves}\label{sec:B}

Here we discuss some properties of the wave motion itself, assuming that there is no slow current, i.e. $\bm V = 0$. First, we consider an irrotational waves in an ideal fluid of infinite depth. Due to Kelvin's theorem the velocity field is always potential, $\bm u = \nabla \phi$, and due to incompressibility condition the velocity potential $\phi$ satisfies the Laplace equation $\nabla^2 \phi = 0$. The pressure $p_u$ can be found from Bernoulli equation \cite{lamb1975hydrodynamics}
\begin{equation}\label{Bernoulli}
    \partial_t \phi + \frac{1}{2}\big(\nabla \phi\big)^2 + \frac{p_u}{\rho} + gz = 0,
\end{equation}
and one has to supplement these equations by the condition of the absence of the wave motion $\nabla \phi \rightarrow 0$ at infinite depth $z \rightarrow -\infty$, and by the boundary conditions posed at the fluid surface $z=h(x,y,t)$.  This is the kinematic boundary condition (\ref{kinematicBC}) and the dynamic boundary condition (\ref{dynamicBC}), which is simply equal to $p_u\vert_{z=h}=0$ in an ideal fluid.

Since the potential satisfies Laplace equation $(\partial_z^2+\hat k^2)\phi=0$ and decreases downward, one finds $\partial_z \phi = \hat k \phi$, where we introduced the wave number operator $\hat k = (-\partial_x^2 - \partial_y^2)^{1/2}$. In the linear approximation, the boundary conditions have a form $\partial_t\phi^{\scriptscriptstyle (1)}\vert_{z=0}+gh=0$ and $\partial_th=\partial_z\phi^{\scriptscriptstyle (1)}\vert_{z=0}$, and therefore one can obtain the dispersion law $\omega^2 = gk$, where $k>0$ is the wave number.

Next, we begin to describe surface waves in a fluid with low kinematic viscosity $\nu$. Besides the change in potential $\phi$ due to modified boundary conditions, the viscosity produces small vortical correction which is nonzero inside the thin viscous sublayer near the fluid surface and is described by the vector stream function $\psi_{\alpha}$. Now the velocity of the wave motion is ${\bm u} = {\bm u}^\phi + {\bm u}^\psi$, where $u^\phi_i = \partial_i \phi$ and $u_\alpha^\psi = \partial_{z} \psi_\alpha$, $u^\psi_z= -\partial_{\alpha} \psi_\alpha$. The imaginary part of the wave frequency describes the decay of surface waves due to the fluid viscosity, $\omega''=-2 \nu k^2$, and this means that the waveform in the linear approximation satisfies the wave equation \cite{lamb1975hydrodynamics}
\begin{equation}\label{wave_equation}
    \partial_t^2 h + g\hat k h + 4\nu \hat k^2 \partial_t h = 0.
\end{equation}
To be self-consistent, all further calculations should be implemented up to the relative accuracy of ${\mathcal O}(\gamma^2)$. Note that the same equation (\ref{wave_equation}) is valid for the velocity potential $\phi^{\scriptscriptstyle (1)}$.

The vortical part of the velocity field ${\bm u}^\psi$ is small in viscosity, so it should be neglected for the normal component of the dynamic boundary condition $p_u-2\rho \nu n_in_j\partial_iu_j=0$, see equation (\ref{dynamicBC}). Using this condition, the wave equation (\ref{wave_equation}) and the relation $p_u^{\scriptscriptstyle (1)} = - \rho g z - \rho \partial_t \phi^{\scriptscriptstyle (1)}$, which is found from the linearized Bernoulli equation (\ref{Bernoulli}) that is valid in the viscous fluid as well, see Ref.~\cite[\S 349]{lamb1975hydrodynamics}, we conclude that the velocity potential in the linear approximation is equal to
\begin{equation}\label{flow_linear}
    \phi^{\scriptscriptstyle (1)}
    =
    \frac{\partial_t+2\nu\hat k^2}{\hat k}\exp(z\hat k)h.
\end{equation}
Considering the solenoidal part ${\bm u}^\psi$, we should take into account that the wave amplitude $h$ can be larger than the thickness of the viscous sublayer $\delta = \sqrt{2} \gamma/k$. This situation was studied systematically, e.g., in Ref.~\cite{longuet1953mass}. In this case, the viscous boundary layer cannot be described by a simple linear theory. For our purposes it is sufficient to note that the vector stream function decreases when moving downward from the surface as $\exp((z-h)/\delta)$ and it is equal to
\begin{equation}\label{psi}
    \psi^{\scriptscriptstyle (1)}_\alpha\vert_{z=h} =  -2\nu \partial_\alpha h
\end{equation}
at the fluid surface, see Ref.~\cite{longuet1992theory}. 

\end{document}